\documentclass[aps,prb,superscriptaddress,notitlepage,twocolumn,longbibliography,10pt]{revtex4-2}

\usepackage{amsmath,amssymb,amsthm,bm}
\usepackage{graphicx}
\usepackage[colorinlistoftodos]{todonotes}
\usepackage[inline]{enumitem}
\definecolor{darkblue}{rgb}{0.1,0.2,0.6}
\definecolor{darkred}{rgb}{0.8,0.1,0.2}
\definecolor{crimson}{RGB}{164,16,52}
\definecolor{darkgreen}{rgb}{0.31,0.62,0.24}
\usepackage[colorlinks=true, allcolors=crimson]{hyperref}
\usepackage{diagbox}
\usepackage{epigraph}
\usepackage{float}
\usepackage{multirow}
\usepackage{tikz}
\usepackage{mathtools}
\usepackage{braket}
\usepackage{soul} 
\usepackage[all]{xy}
\usepackage{lipsum}

\usepackage{bbm}

\newcommand{\Rom}[1]{\uppercase\expandafter{\romannumeral#1}}

\newcommand{\ex}[1]{\left\langle #1 \right\rangle}

\newcommand{\mc}{\mathcal}

\makeatletter
\renewcommand*\env@matrix[1][*\c@MaxMatrixCols c]{%
	\hskip -\arraycolsep
	\let\@ifnextchar\new@ifnextchar
	\array{#1}}
\makeatother

\newtheorem*{claim*}{Claim}

\usepackage[normalem]{ulem}

\definecolor{darkred}{rgb}{0.8,0.1,0.2}

\newcommand*{\ShortSecTitle}[1]{\section{#1}}

\begin{document}
\title{{Spacetime Supersymmetry in the Truncated Lattice Schwinger Model}}
\author{Yanting Cheng}
\affiliation{Institute of Theoretical Physics and Department of Physics, University of Science and Technology Beijing, Beijing 100083, China}
\affiliation{Department of Physics and Hong Kong Institute of Quantum Science and Technology,
The University of Hong Kong, Pokfulam Road, Hong Kong SAR, China}
\author{Shang Liu}
\email{sliu.phys@gmail.com}
\affiliation{Institute of Physics,
Chinese Academy of Sciences, Beijing 100910, China}
\affiliation{Department of Physics, California Institute of Technology, Pasadena, California 91125, USA}
\affiliation{Kavli Institute for Theoretical Physics, University of California, Santa Barbara, California 93106, USA}

\begin{abstract}
Gauge theories in (1+1)D have attracted renewed attention partially due to their experimental realizations in quantum simulation platforms. In this work, we revisit the {truncated} lattice massive Schwinger model and the {truncated} lattice Abelian-Higgs model in (1+1)D, where to facilitate quantum simulation, the electric field eigenvalues are truncated to a finite subset while preserving the exact gauge and global symmetries. We uncover previously overlooked universal features in these models, including the emergence of a supersymmetric quantum critical point when the Maxwell term's coefficient changes sign. Our primary focus is the truncated lattice Schwinger model at $\theta=0$, a model not equivalent to familiar spin models. We find that upon reversing the sign of the Maxwell term, the second-order {charge conjugation symmetry breaking transition (or confinement-deconfinement transition in a sense)} can become first-order. {Furthermore}, the two types of transitions are connected by a supersymmetric critical point in the tricritical Ising universality class. In the case of truncated Abelian-Higgs model at $\theta=0$, which we find to be equivalent to the quantum Blume-Capel model, the very existence of a {symmetry-breaking} phase requires a negative-sign Maxwell term. Similarly, there is a tricritical Ising point separating first-order and second-order phase transitions. 
\end{abstract}

\maketitle
\ShortSecTitle{Introduction}
Gauge theory stands as a cornerstone of modern theoretical physics, providing a mathematical framework to describe fundamental forces and emergent phenomena \cite{Polyakov1987,Tong2018Gauge}. Its principles play fundamental roles in the understanding of (de)confinement, quantum anomalies, and topological order in exotic quantum matter. Recently, the quantum simulation of gauge theories in one spatial dimension has emerged as a promising tool, offering a manageable setting for investigating complex dynamics and providing deeper insights into the fundamental laws of physics\cite{Peter2005prl,Yoshihisa2006pra,Benni2012prl,Peter2012prl,Wiese2013,Peter2013prx,Peter2014prl,Peter2016prx,Martinez2016Nature,Klco2018PRA,Kokail2019Nature,Surace2020PRX,Yangbing2020,Alexander2020Science,Yuan2022science,Nguyen2022PRXQuantum,Mueller2023PRXQuantum,Pomarico2023Entropy,USTC-thermalization,Zhang2024NP,Charles2024PRE,Farrell2024PRXQ,Mildenberger2025NP}.

In (1+1)D, the Abelian-Higgs model and the (massive) Schwinger model are among the simplest gauge theories, where the electromagnetic field is coupled to one flavor of complex boson field and one flavor of Dirac fermion field, respectively\cite{Anderson1963pr,Brout1964prl,Higgs1964prl,Kibble1964prl,Schwinger1,Schwinger2}. On top of the choices of matter field, there is also a parameter known as the topological angle $\theta$ which may be set to $0$ or $\pi$; other $\theta$-angles are less interesting due to the absence of the charge conjugation symmetry and hence the absence distinct phases. 
Field theoretical properties of these four models can be found, for example, in Refs.\,\onlinecite{Tong2018Gauge,Klebanov2022SchwingerModel}. 

In order to realize those gauge theories in quantum simulation platforms, it is preferable to have lattice versions of the models with finite-dimensional local Hilbert spaces. One common strategy towards this goal is to truncate the possible eigenvalues of electric field strength to a finite subset, in a way that preserves the exact gauge and global symmetries\cite{Horn1981PLB,Orland990NPB,QLM}. If we consider the truncated lattice models corresponding to the four gauge theories mentioned above, it turns out that three out of the four models are equivalent to familiar spin models as summarized in Table \ref{tab:EffectiveSpinModels}. Such equivalence can be established upon ``integrating out'' the matter degrees of freedom using the gauge redundancy: The case of Schwinger model at $\theta=\pi$ is discussed in Refs.\,\onlinecite{Surace2020PRX,Cheng2022Schwinger,Cheng2024NRP,Jad2025review}, and the cases of Abelian-Higgs model are explained in Appendix \ref{app:IntOut}. 

In this work, we focus on the remaining case, truncated lattice Schwinger model at $\theta=0$, which is not equivalent to any familiar spin model. We determine the phase diagram and low-energy properties of this lattice model. Curiously, we go beyond the conventional parameter regime and allow the coefficient of Maxwell term (electric field energy term) to change sign, leading to surprisingly interesting results. When the coefficient is positive, there is a continuous {charge conjugation symmetry-breaking} phase transition in the Ising universality class upon increasing the fermion mass. The same happens in the untruncated continuum version of the theory. 
When the coefficient is negative, it turns out that the phase transition can become first-order and the two types of phase transition lines are connected by a tricritical point. The tricritical point is in the tricritial Ising universality class which is well-known to have emergent spacetime supersymmetry. {In fact, in a sense to be made clear later, the charge conjugation symmetry breaking transition may be regarded as a confinement-deconfinement transition. We will explain this perspective in some detail and discuss its physical implications. }

We have also found similar interesting phenomena in the truncated lattice Abelian-Higgs model at $\theta=0$. With a positive Maxwell term, the model is always in the {disordered} phase and has no phase transition. However, a negative Maxwell term leads to {an ordered} phase. The transition between these phases includes both a first-order part and an Ising-type second-order part, separated by a tricritical Ising point. These follow from known properties of the quantum Blume-Capel model as explained in Appendix \ref{app:IntOut}. We will not further discuss this model in the rest of the main text.

\begin{table*}[t]
    \centering
    \begin{tabular}{|c|c|c|}
        \hline
        &Abelian-Higgs model &Massive Schwinger Model\\
        \hline
       $\theta=0$ & Quantum Blume-Capel Model & \textit{Spin-1 Model in This Work}\\
        \hline
        $\theta=\pi$& Quantum Ising Model& PXP Model\\
        \hline
    \end{tabular}
        \caption{Equivalent spin models of truncated (1+1)D lattice gauge theories.}
        \label{tab:EffectiveSpinModels}
    \end{table*}

\ShortSecTitle{Model}
\begin{figure}
    \centering   \includegraphics[scale=0.15]{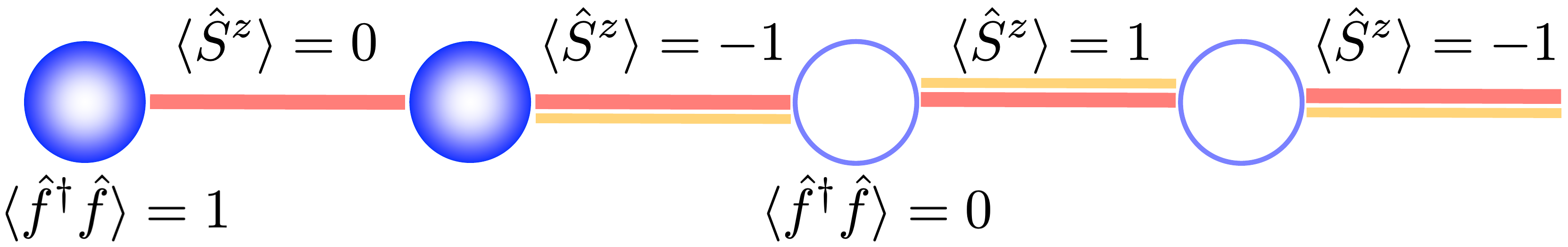}
    \caption{A schematic diagram of a possible configuration. 
    }
    \label{fig:schematic}
\end{figure}
The Kogut-Susskind Hamiltonian of the massive Schwinger model on a lattice is given by \cite{lattice_hamiltonian}
    \begin{eqnarray}
        \hat{H}&=&\frac{1}{2a}\sum_n[\hat{f}^\dagger_ne^{i\hat{\phi}(n)}\hat{f}_{n+1}+h.c.]+m\sum_n(-1)^n\hat{f}^\dagger_n\hat{f}_n\nonumber\\
        &+&\frac{1}{2}g^2a\sum_n(\hat{L}(n)-\frac{\theta}{2\pi})^2,
        \label{Schwinger_Hamiltonian}
    \end{eqnarray}
    where $a$ is the lattice constant, $g$ is the elementary charge, and $m$ is the mass of the fermion. The operators $\hat{f}_n^\dagger$ and $\hat{f}_n$ represent the creation and the annihilation operator of a fermion on the site $n$, and the gauge field operators living on the link between site $n$ and site $n+1$ obey the commutation relation
    \begin{eqnarray}
        [\hat{\phi}(n),\hat{L}(l)]=i\delta_{n,l},
    \end{eqnarray}
which implies $[\hat{L}(n),e^{i\hat{\phi}(n)}]=e^{i\hat{\phi}(n)}$. The eigenvalues of $\hat\phi$ are defined modulo $2\pi$, namely $\phi\sim\phi+2\pi$. Accordingly, the eigenvalues of $\hat L$ are integers $\mathbb{Z}$. The electric field strength is defined as $\hat{E}(n)=\hat{L}(n)-\frac{\theta}{2\pi}$, where $\theta$ angle is a background electric field. The Hamiltonian (\ref{Schwinger_Hamiltonian}) is invariant under the local U(1) transformation that {$\hat{f}_{n}\rightarrow\hat{f}_{n}e^{i\alpha(n)}$,  $e^{i\hat{\phi}(n-1)}\rightarrow e^{i(\hat{\phi}(n-1)-\alpha(n))}$ and $e^{i\hat{\phi}(n)}\rightarrow e^{i(\hat{\phi}(n)+\alpha(n))}$} \emph{for any particular $n$}. Such gauge transformations are generated by the Gauss's law operator:  
    \begin{eqnarray} 
    \hat{G}(n)=\hat{L}(n)-\hat{L}(n-1)-\hat{f}^\dagger_n\hat{f}_n+\frac{1}{2}[1-(-1)^n]. \label{Schwinger_G}
    \end{eqnarray} 
    If not otherwise specified, we focus on the gauge invariant subspace defined by $\hat G(n)=0$ for all $n$. 
    
    In the rest of this work, we only consider $\theta=0$. The Hamiltonian (\ref{Schwinger_Hamiltonian}) then exhibits the following charge conjugation symmetry: 
    \begin{eqnarray}
        &\mathcal{C} e^{i\hat{\phi}(n)}\mathcal{C}^{-1}=e^{-i\hat{\phi}(n+1)},\quad \mathcal{C}\hat{E}(n)\mathcal{C}^{-1}=-\hat{E}(n+1)\nonumber\\
        &\mathcal{C}\hat{f}^\dagger_n\mathcal{C}^{-1}=\hat{f}_{n+1},\quad \mathcal{C}\hat{f}_n\mathcal{C}^{-1}=\hat{f}^\dagger_{n+1}. 
    \end{eqnarray}
    If all low-energy states are nearly invariant under the action of $\mc C^2$, we may expect $\mc C^2$ to approach the identity operator in the continuum limit, in which case the charge conjugation symmetry effectively becomes a $\mathbb{Z}_2$ symmetry. 
    
    From the perspective of experimental realization, it is preferable to have finite-dimensional local Hilbert spaces. This can be done by truncating the allowed eigenvalues of the electric field $\hat E(n)$. We choose the simplest nontrivial truncation where $\hat E(n)$ can take values from $0$ and $\pm1$, implying the following substitution: $\hat{L}(n)\rightarrow\hat{S}_{n,n+1}^z$ and $e^{i\hat{\phi}(n)}\rightarrow\hat{S}_{n,n+1}^+/\sqrt{2}$. Moreover, we also implement the following transformations:  
    \begin{align}
    \begin{cases}
        \text{Odd $n$}:& f_n\rightarrow f_n^\dagger,~ f_n^\dagger\rightarrow f_n, \\
        \text{Even $n$}:& \hat{S}_{n,n+1}^z\rightarrow-\hat{S}_{n,n+1}^z,~ \hat{S}_{n,n+1}^\pm\rightarrow\hat{S}_{n,n+1}^\mp.  
        \end{cases}
        \label{eq:HalfPHTransf}
    \end{align}
which will help simplify the charge conjugation symmetry. 
The resulting Hamiltonian is given by 
    \begin{eqnarray} 
    \hat{H}_{\rm trunc}&=\frac{1}{2\sqrt{2}a}\sum_n[\hat{f}_n\hat{S}_{n,n+1}^+\hat{f}_{n+1}+h.c.]+m\sum_n\hat{f}^\dagger_n\hat{f}_n\nonumber\\
    &+\frac{1}{2}g^2a\sum_n(\hat{S}_{n,n+1}^z)^2. 
    \end{eqnarray} 
    This truncated lattice Schwinger model upholds an exact local U(1) gauge symmetry as well as a charge conjugation symmetry, similar to the original Schwinger model, although we truncated the values of the electric field. The Gauss’s law operator is now given by 
    \begin{eqnarray}        \hat{\tilde{G}}(n)=(-1)^{n+1}(\hat{S}_{n,n+1}^z+\hat{S}_{n-1,n}^z+\hat{f}^\dagger_n\hat{f}_n). 
    \label{eq:TildeGn}
    \end{eqnarray}
Within the $\hat{\tilde G}(n)=0$ subspace, the charge conjugation symmetry of $\hat H_{\rm trunc}$ is simply the translation symmetry. 
In terms of the truncated electric field $\hat{E}_{n,n+1}=(-1)^{n+1}\hat{S}_{n,n+1}^z$, we can rewrite the Gauss's law as $\hat{E}_{n,n+1}-\hat{E}_{n-1,n}=\hat{\tilde{G}}(n)+(-1)^n\hat{f}_n^\dagger\hat{f}_n$. We will refer to $\hat{\tilde{G}}(n)$ as the (nondynamical) gauge charge and $(-1)^n\hat{f}_n^\dagger\hat{f}_n$ as the physical charge. Fig. \ref{fig:schematic} exemplifies a situation with no gauge charge present for the periodic boundary condition. 

The gauge constraints indicate that the degrees of freedom associated with the matter field are entirely redundant. Especially, once the values of $\hat{S}^z_{n,n+1}$ for all links are known, the values of $\hat{f}^\dagger_n\hat{f}_n$ can be derived by Gauss's law. This observation implies that the model can be further simplified by ``integrating out'' the matter sites using the gauge redundancy. 
If $\hat{\tilde G}(n)=0$, the resulting Hamiltonian is 
    \begin{align}
        &\hat{H}_{\textrm{spin-1}}=\nonumber\\
        &\sum_n\left[\frac{\sqrt{2}}{2a}\hat{S}_{n,n+1}^x-2m\hat{S}^z_{n,n+1}+\frac{1}{2}g^2a(\hat{S}_{n,n+1}^z)^2\right], 
    \end{align}
    under the constraint that  $\hat{S}_{n,n+1}^z+\hat{S}_{n-1,n}^z= - \hat{f}_n^\dagger\hat{f}_n$ can only take eigenvalues $0$ and $-1$. Nonzero values of $\hat{\tilde G }(n)$ will not affect the Hamiltonian up to an additive constant, but will modify the Hilbert space constraints as can be derived from \eqref{eq:TildeGn}.  This equivalent formulation of the truncated Schwinger model obscures the gauge symmetry structure, but makes it easier to analyze the phase diagram. 

\ShortSecTitle{Phase Diagram}
\begin{figure}
    \centering
    \includegraphics[scale=0.8333]{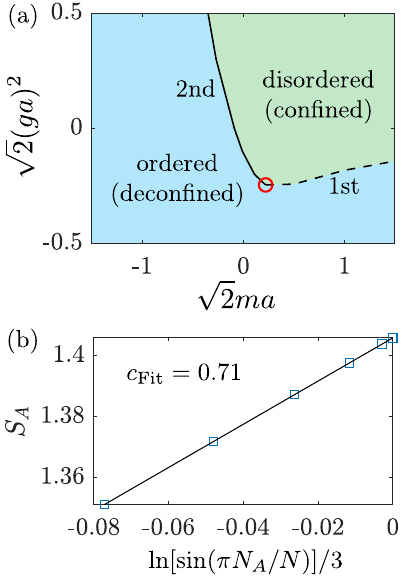}
    \caption{(a) Phase diagram of the spin-1 massive lattice Schwinger model. A dashed line signifies the first-order phase transition and a solid line represents the second-order phase transition. The tricritical point, where these two lines intersect, is marked by a red circle. (b) Central charge at the tricritical point. {Here we fix the system size $N=24$ and compute the ground state entanglement entropy $S_A$ of an interval with a varying length $N_A$. The central charge $c$ can then be extracted from a linear fitting according to the Calabrese-Cardy formula $S_A=c\ln[\sin(\pi N_A/N)]/3+S_0(N)$} \cite{CCFormula}.} 
    \label{fig:PhaseDiagram}
\end{figure}
Recall we focus on $\theta=0$ where the Schwinger model has a global charge conjugation symmetry. In $\hat H_{\rm trunc}$ or $\hat H_{\rm spin-1}$, this is nothing but the one-site translation symmetry. We denote the generator of this symmetry by $\hat{M}_1$. For instance, $\hat{M}_1\hat{A}_{n,n+1}\hat{M}_1^{-1}=\hat{A}_{n+1,n+2}$. One can check that when $\hat{\tilde G}(n)=0$, $\hat M_1$ respects the gauge constraints and commutes with the Hamiltonian. 
We assume that in the continuum limit, this translation symmetry becomes an emergent $\mathbb{Z}_2$ symmetry, i.e. two-site translation acts trivially. Indeed, we have found no spontaneous breaking of the two-site translation symmetry in the parameter regime we explored, which is a necessary condition of the above assumption. 
In line with this emergent $\mathbb{Z}_2$ symmetry, we introduce the order parameter $\hat{O}_n=(-1)^{n}(\hat{S}_{n,n+1}^z-\hat{S}_{n-1,n}^z)$ to differentiate between the ordered and disordered phases as depicted in Fig. \ref{fig:PhaseDiagram} (a). In the ordered phase where the $\mathbb{Z}_2$ symmetry is broken, $\lim_{r\rightarrow\infty}\braket{\hat{O}_n\hat{O}_{n+r}}\neq0$. Conversely, in the disordered phase, $\lim_{r\rightarrow\infty}\braket{\hat{O}_n\hat{O}_{n+r}}=0$.

    Fig. \ref{fig:PhaseDiagram}(a) is our phase diagram from numerical exact diagonalization, where both a first-order (dashed) and a second-order (solid) phase transition lines exist. 
    {We can distinguish first- and second-order phase transitions by extracting the central charge $c$ from entanglement entropy \cite{CCFormula}. At a second-order phase transition point described by a unitary conformal field theory (CFT), we should have $c\geq 0.5$, while at a first-order phase transition point or in any gapped phase, we have $c=0$. Details of our numerical method and the results on central charge can be found in Appendix \ref{app:Numerics}. }
    

The first-order and second-order phase transition lines meet at a tricritical point marked by the red circle in the phase diagram. We anticipated that this tricritical point is in the tricritical Ising universality class with central charge $c=7/10$, and thus determined its precise location, $\sqrt{2}(ma,g^2a^2)=(0.224,-0.246)$ for $N=24$, by maximizing the central charge fitted from entanglement entropy. See Fig. \ref{fig:PhaseDiagram} (b) for the central charge fitting right at this tricritical point. The result is indeed close to the expected theoretical value. 

\begin{figure}
    \centering
    \includegraphics[width=1\linewidth]{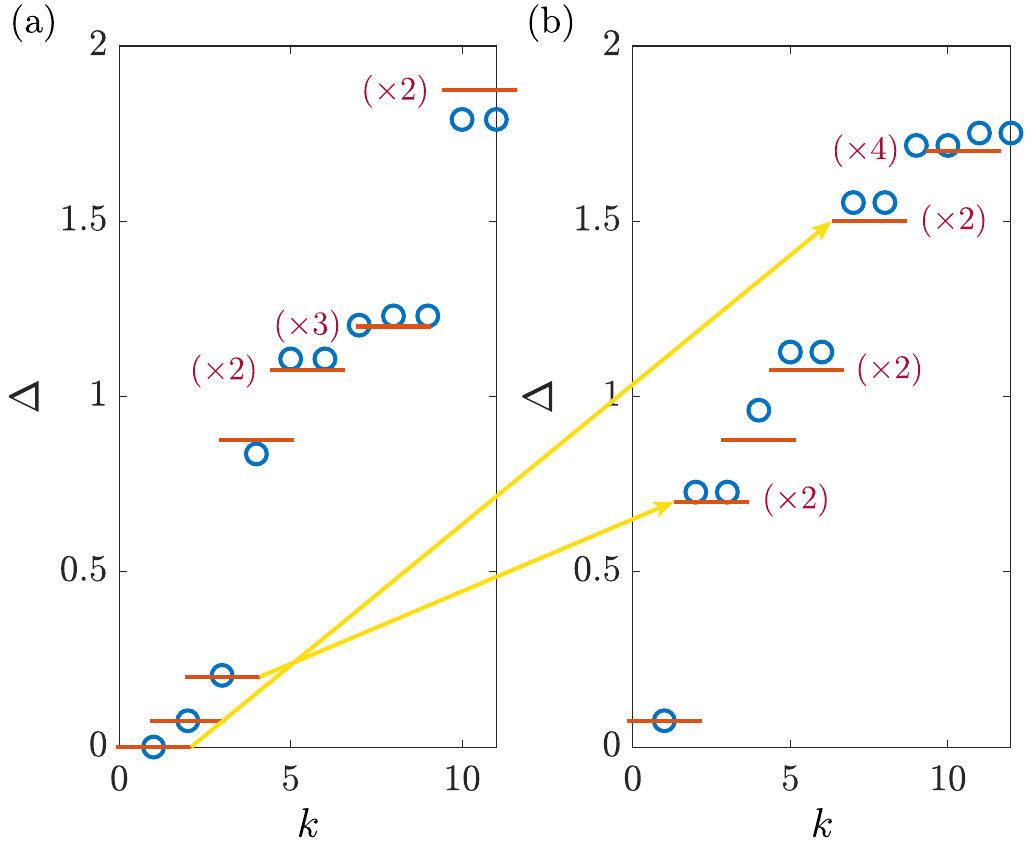}
    \caption{Low energy spectra for (a) $N=28$ and (b) $N=27$. They correspond to the periodic and antiperiodic boundary conditions of the tricritical Ising field theory, respectively. }
    \label{fig:Spectra}
\end{figure}

The CFT describing the tricritical Ising point is known to have spacetime supersymmetry, which imposes nontrivial relations between the energy spectra of periodic and antiperiodic boundary conditions (PBC and APBC). Therefore, let us compare the low-lying energy spectra of our lattice model with those predicted by the tricritical Ising conformal field theory. Note that in our lattice model, the Ising $\mathbb{Z}_2$ symmetry is realized as the translation symmetry, hence PBC and APBC of the tricritical Ising theory should correspond to system size $N$ being even and odd, respectively. In general, conformal symmetry in (1+1)D implies the energy eigenvalues $E_k$ to take the following form: 
\begin{align}
    E_k=\epsilon_0(N)+\frac{2\pi v}{N a}\Delta_k. 
\end{align}
Here, $k=1,2,\cdots$ labels the energy eigenstates, $\epsilon_0$ is a system size dependent energy shift, $v$ is the ``speed of light'', and $\Delta_k$ are a set of universal numbers that depend on the specific conformal field theory and on the boundary condition. The set of $\Delta_k$ for the tricritical Ising theory with both PBC and APBC can be found, e.g. in Ref.\,\onlinecite{Zou2020SCFT}. In Fig.\,\ref{fig:Spectra}, we have shown the comparison between the theoretical and numerical spectra. The two lowest states with $N=28$ and one lowest state with $N=27$ have been used to fix the three constants: $v,\epsilon_0(28), \epsilon_0(27)$. One can see that the remaining energy eigenvalues and degeneracies all fit reasonably well. As mentioned above, the spacetime supersymmetry relates PBC and APBC energy spectra. For example, a pair of superconformal symmetry generators (level $-1/2$) map the PBC state with $\Delta =1/5$ to a pair of APBC states with $\Delta=1/5+1/2=7/10$.  Another pair of superconformal symmetry generators (level $-3/2$) map the PBC ground state with $\Delta=0$ to a pair of APBC states with $\Delta = 0+3/2=3/2$. These and more other examples are all manifest in Fig.\,\ref{fig:Spectra}, confirming the emergence of supersymmetry.

\ShortSecTitle{Confinement-deconfinement transition}
The spontaneous breaking of charge conjugation symmetry implies the deconfinement of charged particles and vice versa. 
We will first give a physical argument of this statement \footnote{{See also the discussions about the (1+1)D Abelian-Higgs model in Ref.\,\onlinecite{Tong2018Gauge}. }}, and then numerically test it in our truncated lattice Schwinger model. 

\begin{figure}
    \centering
    \includegraphics{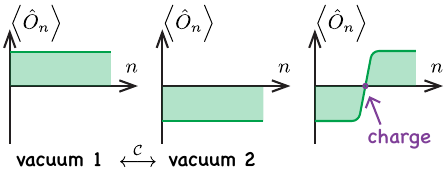}
    \caption{A domain wall between degenerate vacua related by charge conjugation symmetry carries a nonzero charge. }
    \label{fig:DomainWallCharge}
\end{figure}
To facilitate our discussion, it is convenient to smear the notion of charge by defining $\hat q_{n,n+1}:=(\hat E_{n+1,n+2}-\hat E_{n-1,n})/2$, which is the average charge over two neighboring sites $n$ and $n+1$. Given the unbroken two-site translation symmetry, we have $\ex{\hat q_{n,n+1}}=0$ in any vacuum state throughout the phase diagram. In the following, we will only consider such smeared charges if not otherwise specified. We say the system is in the deconfined (confined) phase if we are able (unable) to create isolated charges without costing extensive energy. {Note that this definition of (de)confinement is different from another common definition in the literature based on 1-form symmetry breaking.} 

Suppose the charge conjugation symmetry is spontaneously broken. We must have at least two degenerate vacua \footnote{The term ``vacua'' refers to ground states satisfying the cluster decomposition property. } with nonzero and opposite order parameter values; they are related to each other by the action of $\mc C$. Note that the order parameter $\hat O_n$ is actually proportional to the smeared electric field $(\hat E_{n-1,n}+\hat E_{n,n+1})/2$. Consequently, a domain wall between the two vacua carries a nonzero charge -- proportional to the difference of order parameter values from the two vacua. See Fig.\,\ref{fig:DomainWallCharge} for an illustration. Such a domain wall can be created by applying the charge conjugation on half of the space of one vacuum state. 
The existence of an unscreened charge with a localized energy density implies deconfinement. Conversely, if the system is in the deconfined phase, we should be able to create a pair of unscreened charges out of one vacuum and separate them apart, generating an energy density that is only nonzero near the charges. Now if we send the two charges all the way to spatial infinity, we have created a new vacuum state whose order parameter value differs from the original one. Therefore, at least one of the two vacua has a nonzero order parameter, implying the spontaneous breaking of the charge conjugation symmetry. 

\begin{figure}
    \centering
    \includegraphics[scale=0.5]{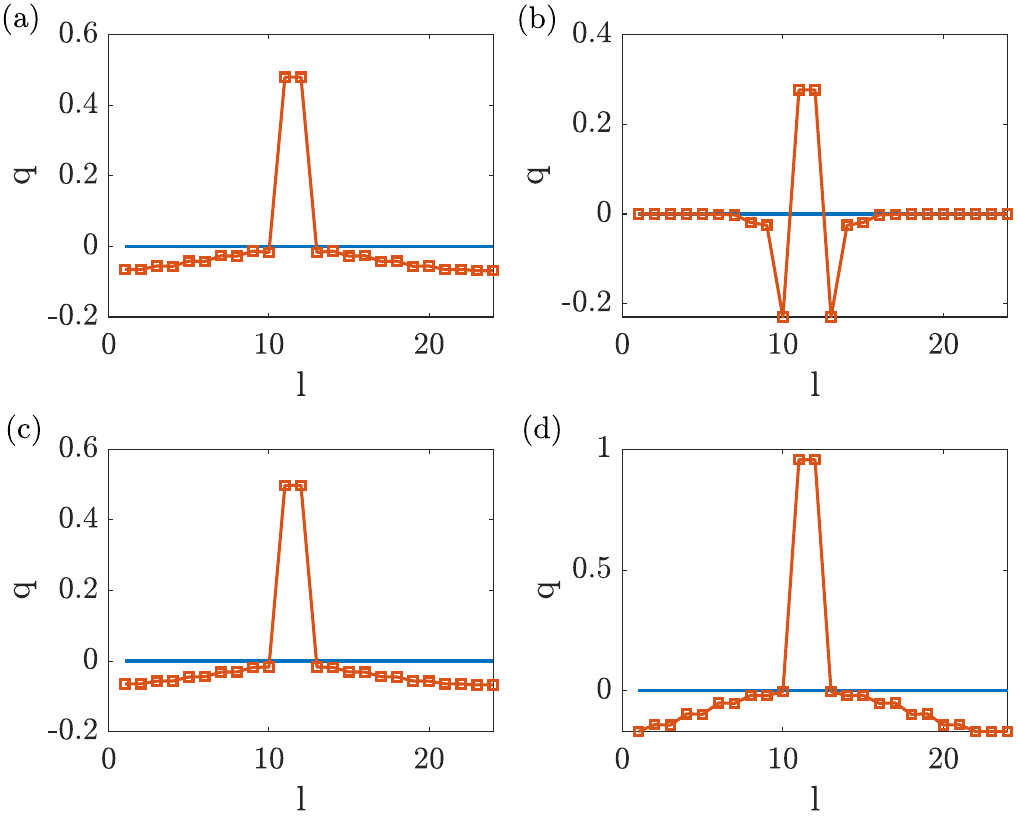}
    \caption{The smeared charge distribution after adding gauge charges into the system for different phases. ($\sqrt{2}ma, \sqrt{2}(ga)^2)=(-1,0.5)$ (a), $(1,0.5)$ (b), $(-1,-0.5)$ (c), $(1,-0.5)$ (d).}
    \label{fig:confinement-deconfinement}
\end{figure}

To numerically test whether a phase is confined or not, we introduce a localized gauge charge into the system and calculate the smeared charge distribution in the new ground state \cite{Wei2024PRR}. In the confined phase, all gauge charges must be fully screened. Equivalently, if there exists some unscreened gauge charge, the system must be in the deconfined phase. 
Our numerics are indeed consistent with the identification between ordering/disordering and deconfinement/confinement. A few examples are shown in Fig. \ref{fig:confinement-deconfinement}, where we modify $\hat{\tilde G}(n)$ at the middle site $n=12$ to be $+1$ for panels (a)-(c) and $+2$ in panel (d). 
For reference, the blue line represents the total charge distribution without any gauge charge. In Fig. \ref{fig:confinement-deconfinement} (b), charges are localized around the additional gauge charge introduced, consistent with confinement. Conversely, in panels (a), (c), and (d), charges are delocalized, signifying deconfinement. Interestingly, we note that in the case of Fig. \ref{fig:confinement-deconfinement} (d), while two units of gauge charges are clearly unscreened as shown in the figure, one unit of gauge charge appears to be fully screened, unlike in panels (a) and (c). This suggests that in the same \emph{bulk} deconfined phase, there exist distinct \emph{defect} phases where we regard the static gauge charge as a (0+1)D defect. The phase transition between them is analogous to the so-called quantum wetting transition \cite{huse1982domain,huse1983melting,campostrini2015quantum1,campostrini2015quantum2,hu2021first,Wang2025RydbergDefect}. {A physical picture and a numerical verification of this defect phase transition are provided in Appendix \ref{app:DefectPT}.} 

\ShortSecTitle{Discussion}
To summarize, we have discovered a supersymmetric quantum critical point -- tricritical Ising point -- in both the truncated lattice Schwinger model and the truncated (1+1)D Abelian-Higgs model. Either case requires a negative-sign Maxwell term. An important theoretical question to explore is whether the aforementioned critical point exists in the untruncated or continuum versions of these models, in which cases a positive $\hat E^4$ term will be needed to bound the energy from below. It would also be interesting to find experimental realizations. Compared to another recent proposal for quantum simulating tricritical Ising point \cite{Li2024SUSY}, our model makes it easier to realize antiperiodic boundary condition, but harder to realize an explicit fermion operator with a Jordan-Wigner string. 

\ShortSecTitle{Acknowledgments}
We thank Hui Zhai, Zhen-Sheng Yuan, Hanteng Wang and Ming-Gen He for the discussion. Y. C. is supported by NSFC Grant No. 12204034, No. 12374251,  Fundamental Research Funds for the Central Universities (No.FRFTP22-101A1), HK CRF C4050-23GF, and C7012-21GF. S. L. acknowledges support from the Gordon and Betty Moore Foundation under Grant No. GBMF8690, the National Science Foundation under Grant No. NSF PHY-1748958, and the Simons Foundation under an award to Xie Chen (Award No.  828078). 

\appendix
\begin{widetext}
\section{Integrating Out Matter Field}\label{app:IntOut}
\subsection{Abelian-Higgs Model}
The (1+1)D Abelian Higgs (AH) model can be defined on a 1D chain with sites labeled by $n\in \mathbb{Z}$ and links labeled by $(n,n+1)$. 
The physical degrees of freedom on each site (link) may be regarded as a particle moving on a ring with the periodic boundary condition (some twisted boundary condition). On a site, we have 
\begin{align}
	[\hat{\varphi},\hat{p}]=i,\quad\varphi\sim\varphi+2\pi,\quad p\in\mathbb{Z}. 
\end{align}
It follows that $e^{i\hat{\varphi}}(\hat{p}+1)=\hat{p}e^{i\hat{\varphi}}$, which means the effect of $e^{i\hat{\varphi}}$ is to increase ${p}$ by $1$. On a link, we have
\begin{align}
	[\hat{a},\hat{E}]=i,\quad a\sim a+2\pi,\quad E\in\mathbb{Z}-\frac{\theta}{2\pi}, 
\end{align}
where the parameter $\theta$ should be identical for all links. Similarly, $e^{i\hat a}$ increases $E$ by $1$. The AH model has the following Hamiltonian. 
\begin{align}
	\hat{H}_{\rm AH}=\sum_n\left( -Je^{i\hat{\varphi}_n}e^{i\hat{a}_{n,n+1}}e^{-i\hat{\varphi}_{n+1}}+h.c. \right)+\frac{1}{2}\mu\sum_n\hat{p}_n^2+\frac{1}{2}g^2\sum_i \hat{E}_{n,n+1}^2. 
\end{align}\\
In addition, the following Gauss' law is imposed on the Hilbert space
\begin{align}
	\hat{G}_n:=\hat{E}_{n,n+1}-\hat{E}_{n-1,n}-\hat{p}_n=0. 
	\label{eq:GaugeConstraint}
\end{align}
Notice that $\hat{G}_n$ commutes with $\hat{H}_{\rm AH}$, thus the physical subspace defined above is an invariant subspace of $\hat{H}_{\rm AH}$. 
We assume $J\geq 0$ without loss of generality and also $\mu\geq 0$. The first two terms are obtained by minimally coupling the O(2) rotor model \cite{Sachdev2011Book} to the U(1) gauge field and the last term is the Maxwell term (gauge field kinetic energy). We expect that in some continuum limit, this lattice model should describe a complex boson field $\phi$ coupled with electromagnetic field. Large $\mu/J$ should correspond to large positive boson mass, while $\mu/J\rightarrow 0$ corresponds to large negative boson mass. However, we did not attempt to carefully derive the continuum limit, thus our convention for the coupling constants is somewhat arbitrary.

The gauge constraint \eqref{eq:GaugeConstraint} implies that the matter field degrees of freedom are completely redundant. Indeed, once we know the values of $E_{n,n+1}$ for all $n$, we can derive the values of all $p_n$ using the Gauss' law. This observation formally means the following isomorphism of Hilbert spaces. 
\begin{align}
	(\text{sites}\otimes\text{links})|_{\hat{G}_n=0~\forall n}&\quad\rightarrow\quad \text{links}\\
	\ket{\{{p}_n\},\{{E}_{n,n+1}\}}&\quad\mapsto\quad \ket{\{{E}_{n,n+1}\}}\nonumber
\end{align}
It is not hard to check that the above explicit mapping is indeed one-to-one. It follows that we can ``integrate out'' all the matter field\footnote{Although we say ``integrate out'', no physical degree of freedom is ignored during this process. }. The effective Hamiltonian thus obtained is 
\begin{align}
	\hat{H}_{\rm eff}&=\sum_n\left( -Je^{i\hat{a}_{n,n+1}}+h.c.\right)+\frac{1}{2}\mu\sum_i(\hat{E}_{n,n+1}-\hat{E}_{n-1,n})^2+\frac{1}{2}g^2\sum_i \hat{E}_{n,n+1}^2
	\nonumber\\
	&=\sum_n\left( -Je^{i\hat{a}_{n,n+1}}+h.c.\right)-\mu\sum_i \hat{E}_{n-1,n}\hat{E}_{n,n+1}+\left(\frac{1}{2}g^2+\mu\right)\sum_i \hat{E}_{n,n+1}^2. 
\end{align}

\subsubsection{Truncated Model at $\theta=\pi$}
Now consider $\theta=\pi$ and we truncate the electric field space to only retain $E=\pm 1/2$. We can then substitute
\begin{align}
	\hat{E}\mapsto\begin{pmatrix}
	1/2 & 0\\
	0 & -1/2
	\end{pmatrix},
	\quad
	e^{i\hat{a}}\mapsto\begin{pmatrix}
	0 & 1\\
	0 & 0
	\end{pmatrix}. 
\end{align}
These substitutions preserve the U(1) gauge symmetry generated by $G_n$ as well as the global charge conjugation symmetry $(\hat{E},\hat{a},\hat{p},\hat{\varphi})\mapsto -(\hat{E},\hat{a},\hat{p},\hat{\varphi})$. 
The procedure of integrating out matter fields works in the same way, and the effective Hamiltonian becomes 
\begin{align}
	\hat{H}_{\rm eff}\mapsto -J\sum_i \hat{X}_{i+1/2}-\frac{1}{4}\mu\sum_i\hat{Z}_{i-1/2}\hat{Z}_{i+1/2}, 
\end{align}
which is exactly the transverse field Ising model. The ordered (disordered) phase of the Ising model corresponds to the deconfined (confined) phase of the AH model. 
\subsubsection{Truncated Model at $\theta=0$}
At $\theta=0$, we may truncate the model to by retaining $E=0,\pm1$, or equivalently applying the substitutions
\begin{align}
	\hat{E}\mapsto\begin{pmatrix}
		1 & 0 & 0\\
		0 & 0 & 0\\
		0 & 0 & -1
	\end{pmatrix},
	\quad
	e^{i\hat{a}}\mapsto\begin{pmatrix}
		0 & 1 & 0\\
		0 & 0 & 1\\
		0 & 0 & 0
	\end{pmatrix}. 
\end{align} 
Using spin-1 operators, the resulting effective Hamiltonian is
\begin{align}
	H_{\rm eff}\mapsto -\sqrt{2} J\sum_n \hat{S}^x_{n,n+1} -\mu\sum_i \hat{S}^z_{n-1,n}\hat{S}^z_{n,n+1}+\left(\frac{1}{2}g^2+\mu\right)\sum_i (\hat{S}^z_{n,n+1})^2. 
\end{align}
This is the quantum Blume-Capel model \cite{Alcaraz1985QuantumBCModel}. The schematic phase diagram looks like Fig.\,\ref{fig:AHTheta0PhaseDiagram}; see Refs.\,\onlinecite{Alcaraz1985QuantumBCModel} and \onlinecite{Xavier2011PreciseCriticalPoint} for details. Note that there is no confinement-deconfinement transition when the Maxwell term is positive, which is consistent with the result of the continuum AH model \cite{Tong2018Gauge}. 

\begin{figure}
	\centering
	\includegraphics{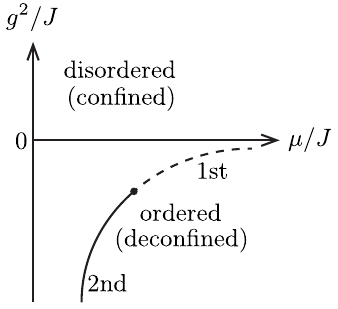}
	\caption{Schematic phase diagram of the truncated lattice AH model at $\theta=0$. }
	\label{fig:AHTheta0PhaseDiagram}
\end{figure}

We note that a very different version of (1+1)D lattice AH model exists in the literature \cite{ChandaAH1,ChandaAH2}, whose phase diagram differs a lot from our model, and contains a novel $c=3/2$ quantum critical point. We refer interested readers to those references for details. 

\subsection{Schwinger Model}
For (truncated) lattice Schwinger models, such as the $\theta=0$ model studied in the main text, the procedure of integrating out matter sites is similar but slightly more tricky. There are two caveats: (1) The fermion number $\hat f^\dagger_n\hat f_n$ can only be $0$ or $1$, which leads to residue constraints on the spins after integrating out matters. (2) We need to be careful about the ordering of fermion operators. 

Let us elaborate this procedure for the model considered in the main text. We fix the values of $\hat{\tilde G}(n)$, say all zero, and let $\mathcal L_{\rm sites+links}$ be the Hilbert space for both sites and links under the Gauss's law constraints. $\hat H_{\rm trunc}$ acts within this Hilbert space. As we mentioned in the main text, the Gauss's law implies that all site variables are actually redundant. More precisely, we expect $\mathcal L_{\rm sites+links}$ to be isomorphic to another Hilbert space $\mathcal L_{\rm links}$ which only consists of the spins on the links and is subject to the following constraints: $(-1)^{n+1}\hat{\tilde G}(n)-\hat{S}_{n,n+1}^z-\hat{S}_{n-1,n}^z$ can only take eigenvalues $0$ and $1$. Let $\hat\nu_n:=\hat f^\dagger_n \hat f_n$. We can explicitly write down an isomorphism $\mathcal L_{\rm sites+links}\rightarrow \mathcal L_{\rm links}$ as follows: 
\begin{align}
    \ket{\{\nu_n \}, \{S^z_{n,n+1}\} }:=(\hat f_N^\dagger)^{\nu_N}(\hat f_{N-1}^\dagger)^{\nu_{N-1}}\cdots (\hat f_1^\dagger)^{\nu_1}\ket{\{S^z_{n,n+1}\} }
    \quad\mapsto\quad \ket{\{S^z_{n,n+1}\} }. 
\end{align}
One can check that under this isomorphism, we have the following operator mapping rules: 
\begin{align}
    \begin{cases}
        \hat f_n \hat S^+_{n,n+1}\hat f_{n+1}\mapsto \hat{\mc P}\hat S^+_{n,n+1}\hat{\mc P} & 1\leq n\leq N-1\\
        \hat f_N \hat S^+_{N,1}\hat f_{1}\mapsto (-1)^{\nu_N+\nu_{N-1}+\cdots +\nu_1 -1}\hat{\mc P}\hat S^+_{N,1}\hat{\mc P} & n=N
    \end{cases}, 
\end{align}
where $\hat{\mc P}$ is the projection operator onto the allowed spin configurations. 
The sign factor in the second line is actually a constant depending on the values of $\hat{\tilde G}(n)$: 
\begin{align}
    \sum_n\nu_n=\sum_n (-1)^{n+1}\tilde G(n)\mod 2. 
\end{align}
The mapping rules for other operators in the Hamiltonian is rather obvious, for example, 
\begin{align}
    \hat f^\dagger_n \hat f_n\mapsto (-1)^{n+1}\hat{\tilde G}(n)-\hat{S}_{n,n+1}^z-\hat{S}_{n-1,n}^z. 
\end{align}
The extra sign factor mentioned above is rather annoying, but we can actually remove it by applying a local spin rotation generated by $\hat S^z_{N,1}$. Such a rotation commutes with $\hat S^z_{n,n+1}$ and thus preserves both the spin configuration constraints and other terms in the Hamiltonian. Finally, we obtain the equivalent Hamiltonian $\hat H_{\rm spin-1}$ given in the main text. 

The case of $\theta=\pi$ works in a similar way and the result can be found in Refs.\,\onlinecite{Surace2020PRX,Cheng2022Schwinger,Cheng2024NRP}. 

\section{Further Details about the Numerics}\label{app:Numerics}
\subsection{Enforcing Local Constraints}
We numerically solve our spin-1 model using exact diagonalization, and in particular the Lanczos algorithm (Matlab built-in function). We fully utilize the local constraints in the model to reduce Hilbert space dimensions, which enables us to access system sizes up to around $N=26$. In this subsection, we will explain the strategy we used to enforce local constraints. 

Suppose we have a spin chain of $N$ sites labeled by $j=1,2,\cdots,N$. Denote by $\mc L_1$ the local Hilbert space of each site and $\mc L_{\rm ext}=(\mc L_1)^{\otimes N}$ the full tensor product Hilbert space. We would like to impose local constraints between nearest-neighbor pairs of sites: $F_{j,j+1}(O_j,O'_{j+1})=0$, where $O_j$ ($O'_{j+1}$) is a local operator acting on the $j$-th ($(j+1)$-th) site and $F_{j,j+1}(\cdot,\cdot)$ is some analytic function. We make a simplifying assumption: Under some canonical choice of basis vectors for $\mc L_1$, all $O_j$ and $O'_{j+1}$ operators (and hence also $F_{j,j+1}(O_j,O'_{j+1})$) are diagonal in the tensor product basis. Therefore, the effect of local constraints is to select a subset of the tensor product basis vectors. This subset spans the physical Hilbert space $\mc L_{\rm phys}$. Our goal is to construct the matrices of local operators projected onto $\mc L_{\rm phys}$. This, for example, will enable us to construct the Hamiltonian and diagonalize it. 

A na\"{i}ve strategy to achieve this goal is to first construct the operator matrix in $\mc L_{\rm ext}$ and then restrict it to $\mc L_{\rm phys}$. This, however, is not very practical because constructing a matrix in $\mc L_{\rm ext}$ is not memory efficient. We instead take an inductive approach which we now elaborate. For the moment, let us assume that we care about open boundary condition, meaning that we would like to impose the constraints $F_{1,2},F_{2,3},\cdots, F_{N-1,N}$. We define
\begin{align}
    \mc L_k:=\{\text{subspace of }{(\mc L_1)^{\otimes k}}\text{ satisfying }{F_{1,2}, F_{2,3},\cdots,F_{k-1,k}=0}\}. 
\end{align}
Our strategy is to construct the following sequence of inclusion maps (each map can be saved as a logical vector in the numerical program): 
\begin{align}
    f_k: \mc L_k\equiv\{\text{subspace of }{\mc L_{k-1}\otimes\mc  L_1}\text{ satisfying }{F_{k-1,k}=0}\}\hookrightarrow \mc L_{k-1}\otimes \mc L_1. 
\end{align}
Note that $f_k$ contains the same information as the projection map $P_k:\mc L_{k-1}\otimes \mc L_1\rightarrow \mc L_{k}$ enforcing the local constraint $F_{k-1,k}=0$. 
Once this is done, we can project any local operator onto $\mc L_N=\mc L_{\rm phys}$ without explicitly involving $\mc L_{\rm ext}$. As an example, consider an operator $R_k$ acting on the $k$-th site. Let $P:\mc L_{\rm ext}\rightarrow \mc L_{\rm phys}$ be the full projection operator onto the physical subspace. We have
\begin{align}
    PR_kP=P_N(\cdots P_{k+1}(P_k({\rm Id}_{\mc L_{k-1}}\otimes R_k)P_k\otimes {\rm Id}_{\mc L_1})P_{k+1}\cdots\otimes {\rm Id}_{\mc L_1})P_N. 
\end{align}
Note in particular that once we have $f_k$ (or equivalently $P_k$), we are already able to project $O_k$ onto $\mc L_k$, which enables us to construct $f_{k+1}$ -- the inductive construction of the inclusion maps. 

The case of periodic boundary condition is essentially the same. We just need one more inclusion map that enforces the constraint $F_{N,1}=0$.

\subsection{Distinguishing First- and Second-Order Phase Transitions}
In this subsection, we provide more details about how we numerically distinguish first- and second-order phase transitions which both exist in our phase diagram. 

A sharp distinction between first- and second-order phase transitions can be made by extracting the central charge $c$ which intuitively counts the number of gapless modes in the system. At a second-order phase transition point described by a unitary CFT, we should have $c\geq 0.5$, while at a first-order phase transition point or in any gapped phase, we have $c=0$. The central charge can be conveniently extracted by computing the entanglement entropy. Consider a single-component subsystem of length $N_A$ and let $N$ be the total system size. The Calabrese-Cardy formula dictates that 
\begin{align}
    S_A=\frac{c}{3}\ln\left[ \frac{N}{\pi}\sin\left(\frac{\pi N_A}{N}\right) \right]+{\rm const}, 
\end{align}
where $S_A$ is the von Neumann entanglement entropy of the subsystem in the ground state and the additive constant has no dependence on $N_A$ or $N$. This suggests at least two ways to extract $c$: 
\begin{enumerate}
    \item Fix $N$, compute $S_A$ as a function of $N_A$, and then linear fit $S_A$ against $\ln[\sin(\pi N_A/N)]$. 
    \item Fix $N_A/N=1/2$, compute $S_A$ as a function of $N$, and then linear fit $S_A$ against $\ln(N)$. 
\end{enumerate}
We implemented both approaches and the results are shown in Fig.\,\ref{fig:CCFit}. Compared to the phase diagram shown in the main text, we clearly see that the central charge is zero in both gapped phases and along the lower-right part of the phase transition line. In contrast, the central charge takes a nonzero value along the upper-left part of the phase transition line. This suggests the coexistence of first- and second-order phase transitions. 

\begin{figure}
    \centering
    \includegraphics[width=0.8\linewidth]{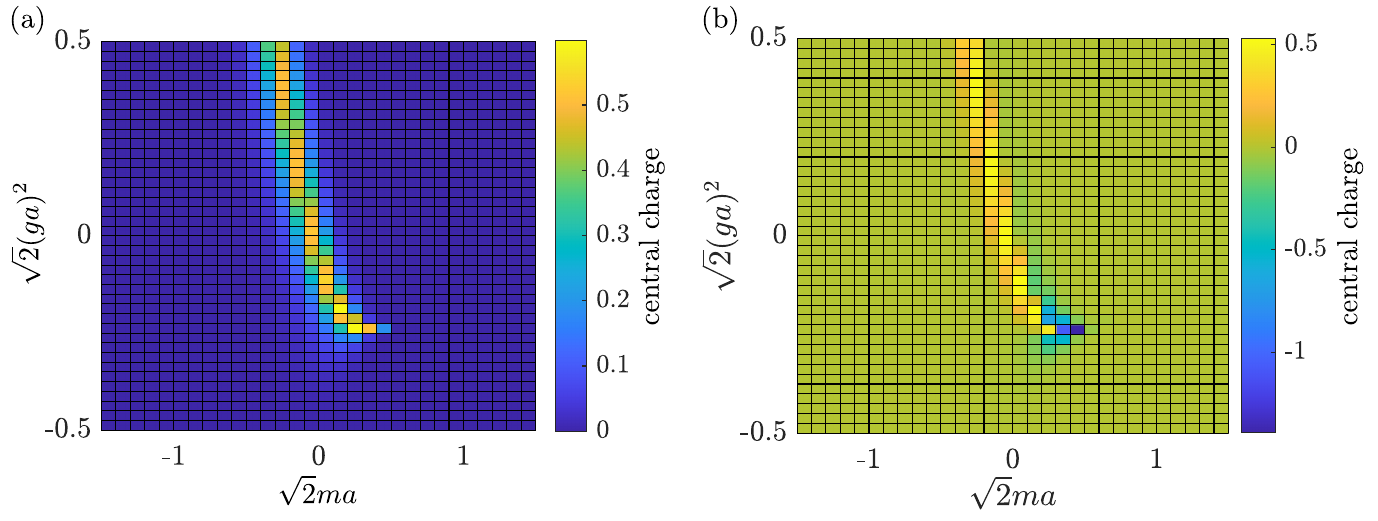}
    \caption{Central charge $c$ extracted from the entanglement entropy by two different approaches: (a) fixing $N=24$ while varying $N_A\in\{7,8,9,10,11,12\}$ and (b) fixing $N_A/N=1/2$ while varying $N\in\{18,20,22,24,26\}$. Note the difference in color bars. }
    \label{fig:CCFit}
\end{figure}

We notice that in panel (b) of Fig.\,\ref{fig:CCFit}, the fitted central charge can become negative, meaning that the entanglement entropy $S_A$ can decrease as $N$ increases. Negative central charge is not actually allowed due to unitarity, and this phenomenon is due to the strong finite-size effect when the correlation length becomes comparable to the system size. 

There is another a technical subtlety in the second approach for extracting central charge (Fig.\,\ref{fig:CCFit}b). Within the ordered phase, the system has two almost degenerate ground states, and as we vary the system size $N$, the numerical algorithm may not consistently converge to the ``same'' state. This can lead to some irregular results in the fitted central charges. To remedy this problem, we always symmetrize the numerically obtained ground state, which in particular means that we choose the symmetric cat states in the order phase. There is no similar issue in the first approach for extracting central charge (Fig.\,\ref{fig:CCFit}a), where we fix a ground state for each set of model parameters and only vary the entanglement cuts.

\section{Defect Phase Transition}\label{app:DefectPT}
We have mentioned in the main text that within the same bulk deconfined phase, there can exist distinct defect phases after adding a static gauge charge. We will elaborate on this point in this section. More specifically, we will first illustrate the physical picture by two limiting cases, and then provide numerical verification for the existence of a defect phase transition. 

As a reminder, the local constraints of the effective spin-1 model read 
\begin{align}
    \hat S^z_{n-1,n}+\hat S^z_{n,n+1}=(-1)^{n+1}\hat{\tilde G}(n)-\hat f^\dagger_n\hat f_n. 
\end{align}
We will choose some even integer $n_0\in 2\mathbb{Z}$ and modify $\hat{\tilde G}(n_0)$ to $+1$. The effect is that for $n=n_0$ and $n\neq n_0$, the allowed eigenvalues of $\hat S^z_{n-1,n}+\hat S^z_{n,n+1}$ are $\{ -1,-2 \}$ and $\{0,-1\}$, respectively. Also recall that the smeared charge $\hat q_{n,n+1}$ is defined as $(-1)^n(\hat S^z_{n+1,n+2}-\hat S^z_{n-1,n})$. 

Consider the following two classical limits. 
\begin{itemize}
    \item $(ga)^2<0$, $ma<0$, $|ma|\gg |ga|^2\gg1$:

    Before adding the gauge charge, the two ground states of the system on a periodic chain with even number of sites are given by
    \begin{align}
        \ket{\Omega_1}=\ket{-0-0-0\cdots},\quad 
        \ket{\Omega_2}=\ket{0-0-0-\cdots}, 
    \end{align}
    where we use $\ket{\pm}$ and $\ket{0}$ to denote the three $S^z$ eigenstates of spin-1. After adding the static gauge charge, whose location we denote by a vertical bar, the ground states are given by
    \begin{align}
        &\ket{-0-0-0|-0-0-0}=\ket{\Omega_1},\\
        &\ket{-0-00-|-0-0-0},\\
        &\ket{-00-0-|-0-0-0},\\
        &\cdots\\
        &\ket{0-0-0-|-00-0-},\\
        &\ket{0-0-0-|0-0-0-}=\ket{\Omega_2}. 
    \end{align}
    We observe that there exists a delocalized charge (the ``00'' configuration). If we treat the $S^x$ term in the Hamiltonian as a perturbation, its effect is to enable the hopping of this delocalized charge. The effective low-energy theory will be a single particle hopping on an \emph{open} chain of length $N/2+1$, with hopping amplitude $ a^{-1}/[4ma+(ga)^2]$. The low-energy spectrum will then be gapless, with excitation energies of the order $1/N^2$. 
    
    \item $(ga)^2<0$, $ma>0$, $|ga|^2\gg1$, $|ma|\gg 1$: 

    Before adding the gauge charge, the two ground states are given by
    \begin{align}
        \ket{\Omega_1'}=\ket{+-+-+-\cdots},\quad\ket{\Omega_2'}:=\ket{-+-+-+\cdots}. 
    \end{align}
    
    After adding the static gauge charge, there are still only two ground states: 
    \begin{align}
        \ket{+-+-+-|0-+-+-},\quad \ket{-+-+-0|-+-+-+}. 
    \end{align}
    Hence the energy spectrum is gapped, and there is no delocalized charge. 
\end{itemize}

\begin{figure}
    \centering
    \includegraphics[width=0.8\linewidth]{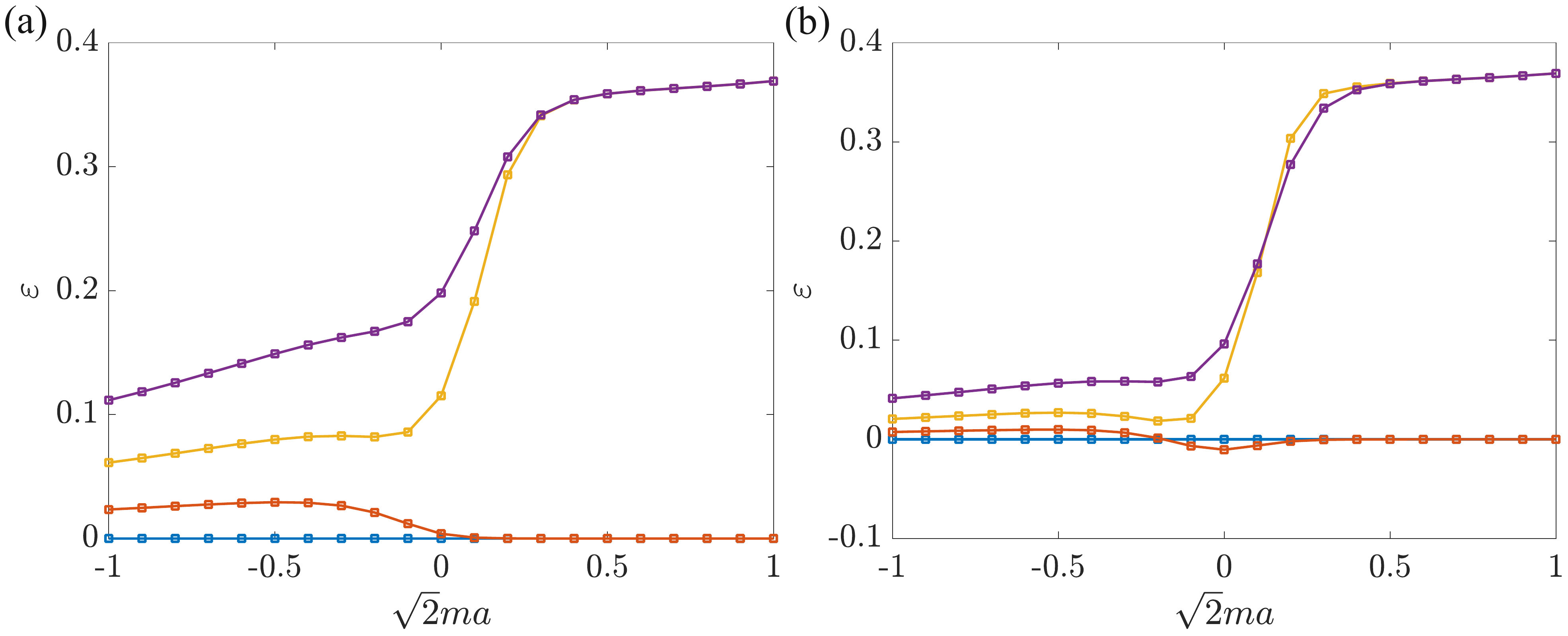}
    \caption{We fix the $\sqrt{2}(ga)^2=-0.5$, and vary the $\sqrt{2}ma$ as the horizontal axis. Then we calculate the four lowest energy levels relative to the ground state. The system size of (a) is 26. We performed finite-size scaling in panel (b), using data from system sizes 18, 20, 22, 24, and 26.}
    \label{fig:wetting}
\end{figure}

From the above analysis, we are expecting the existence of two distinct defect phases (within the bulk deconfined phase) where the static $+1$ gauge charge (at site $n_0\in2\mathbb{Z}$) is unscreened and screened, respectively. In the unscreened phase, the low-energy spectrum for even $N$ should be gapless, while in the screened phase, the spectrum is gapped. To verify this statement numerically, we fix $\sqrt{2}(ga)^2=-0.5$ and compute the low-energy spectrum with defect for a sequence of $\sqrt{2}ma$ ranging from $-1$ towards $+1$. The result of the $4$ lowest energy levels at $N=26$ is shown in Fig.\,\ref{fig:wetting}(a), where we can already see a clear difference between the left and right halves of the plot. We have also performed finite-size scaling analysis, and the result is shown in Fig.\,\ref{fig:wetting}(b). Here, we extrapolate the energy levels to infinite system size with the ansatz $\varepsilon = C_0 + C_1/N^2$ where $\varepsilon$ denotes the energy relative to the ground state. Though still not perfect, Fig.\,\ref{fig:wetting}(b) strongly suggests a transition between gapless and gapped spectra as we anticipated. We leave a thorough analysis of the defect phase transition, including the precise determination of transition points for different values of $(ga)^2$ and the critical exponents, to future works. 
\end{widetext}

\bibliography{Bib_Refs.bib}
\end{document}